\documentstyle[aps,epsf]{revtex}

\newcommand{\beq}{\begin{equation}}
\newcommand{\eeq}{\end{equation}} 

\newcommand{\beqa}{\begin{eqnarray}}
\newcommand{\eeqa}{\end{eqnarray}}

\newcommand{\infig}[2]{\begin{center}\mbox{\epsfxsize #2
\epsfbox{#1}}\end{center}}

\begin{document}

\title{Violation of Bell inequalities by photons more than 10 km apart}
\draft

\twocolumn[\hsize\textwidth\columnwidth\hsize\csname 
@twocolumnfalse\endcsname

\author{W. Tittel, J. Brendel, H. Zbinden and N.
Gisin}

\address{University of Geneva, Group of Applied Physics, 20, Rue de l'Ecole
de M\'edecine, CH-1211 Geneva 4, 
Switzerland\\
e-mail: wolfgang.tittel@physics.unige.ch}
\date{\today} 

\maketitle

\begin{abstract}
A Franson-type test of Bell inequalities by photons 10.9 km apart is
presented. Energy-time entangled photon-pairs are measured
using two-channel analyzers,
leading to a violation of the inequalities by 16 standard deviations without
subtracting 
accidental coincidences. Subtracting them, a 2-photon interference
visibility of 95.5\% is observed, demonstrating that distances up to 10 km
have no significant
effect on entanglement. This sets quantum
cryptography with photon pairs as a practical competitor to the schemes
based on weak pulses.
\end{abstract}

\vskip1pc
]

Quantum theory is nonlocal. Indeed, quantum theory predicts correlations among
distant measurement outcomes that cannot be explained by any theory which
involves
only local variables. This was anticipated by Einstein, Podolski and Rosen
\cite{EPR}
and by Schr\"odinger \cite{Schrodinger}, among others, and first demonstrated
by John Bell in 1964 with his now famous inequality \cite{Bell64}.
However, the nonlocal feature cannot be exploited for superluminal communication
\cite {fasterthanlight}.
Hence, there is no contradiction
with relativity, though there is clearly a tension. Physicists disagree
about the
significance and importance of this tension. This led Abner Shimony to name this
situation "peaceful coexistence between quantum mechanics and relativity"
\cite{Shimony}.

Why should one still bother about quantum nonlocality despite that 
all experiments so far are in agreement with quantum theory
\cite{Bellexp,Tapster94}?
The traditional motivations are based on fundamental
questions on the meaning and compatibility of our basic theories, quantum
mechanics
and relativity: to date, no experiment to test Bell's inequality has been loophole free
\cite{loophole,locality,closeloophole}
and no experiment so far has directly probed the tension
between quantum non locality and relativity. 
Recently, additional motivations to investigate quantum non-locality 
arose based on the potential applications of the fascinating
field of quantum information processing:
all of quantum computation and communication is based on the assumption
that quantum systems can be entangled and that the entanglement can be
maintained over long times and distances \cite{PhysWorld}.

In 1997 we have demonstrated that two-photon correlations remain strong enough over
10 km so that a violation of Bell inequalities could be
expected \cite{Tittel98}. In this letter we report on a new experiment
using two-channel analyzers in which all 4 coincidence rates have been measured simultaneously, 
thus allowing to obtain directly the correlation 
coefficient that defines the Bell inequalities. Our experiment demonstrates
a violation of Bell's inequalities with photons more 
than 10 km apart \cite{notthesame}, even without subtracting the accidental 
coincidences. 
In addition, an experiment with three interferometers, two on one end and the third at the other end 
(10 km away) is presented. The two nearby interferometers analyse the incoming
photons randomly, the choice being made by a passive beam splitter. This setup enables to test directly 
the CHSH form of Bell-inequalities \cite{CHSH}.
Our experiment establishes also the feasibility of quantum cryptography
with photon pairs \cite{Ekert91} (in opposition to weak coherence pulses) over a
significant distance.

For our Franson-type test of Bell inequalities \cite{Franson89}, we produce 
energy-time entangled photons by parametric downconversion (Fig. 1). 
Light from a semiconductor laser with an external cavity (10 mW at 655 nm, $\Delta\nu<$10MHz) 
passes through a dispersion prism P to separate out the 
residual infrared fluorescence light and is focused into a 
$\mbox{KNbO}_3$ crystal. The crystal is oriented to ensure 
degenerate collinear type I phasematching for 
signal and idler photons at 1310 nm \cite{1310}. Behind the crystal, the pump light is separated 
out by a filter F (RG~1000) while the passing down-converted 
photons are focused (lens L) into one input port of a standard 3-dB fiber coupler. Therefore half of the 
pairs are split and exit the source by different output fibers.
Using a telecommunications fiber 
network, the photons are then analysed by 
all-fiber interferometers located 10.9~km apart from one another in the small villages of Bellevue 
and Bernex, respectively. 
The source, located in Geneva, was 4.5 km away from the first analyser and 7.3 km from the
second, with connecting
fibers of 8.1 and 9.3 km length, resp., as indicated in Fig. 1. 
Our interferometers use both the Michelson configuration and have a long
and a short arm. In order to compensate all birefringence effects in the arms (i.e. to 
stabilize the polarization), we employ 
so called Faraday mirrors (FM) to reflect the light \cite{faradaymirror}. 
At the input ports, we use optical circulators (C). These devices
guide the light from the source to the interferometer, but, thanks to the
non-reciprocal
nature of the Faraday effect, guide the light reflected back from the
interferometer to another fiber, serving as second output port. The output ports of each interferometer are
connected to
photon counters \cite{detector}. We
label the "direct" port as "+", the one connected to the circulator "-". 
To control and change the phases ($\delta_1, \delta_2$), the 
temperature of the interferometers can be varied. 

Since the
arm length difference is five orders of magnitude larger than the single photon coherence length, 
there is 
no single photon interference. However, the path difference in both
interferometers is
precisely the same, with a sub-wavelengths accuracy. 
Moreover, this imbalance is two orders of magnitude smaller than the
coherence length of the pump laser. 
Hence, an entangled state can be produced where either both photons
pass through the
short arms or both use the long arms. Noninterfering possibilities (the photons 
pass through different arms)
can be discarded using a high resolution coincidence technique \cite{Brendel91}.

To ensure symmetry for the two channels of each analyzer,
we adjusted the count rates 
of the detectors attached to the same interferometer. Typical rates are 39.5 kHz including 26 kHz dark 
count rates. 
The classical signals from the photon detectors are transmitted back to Geneva.
We measure the four different 
numbers of time-correlated events $R_{i,j}(\delta_1,\delta_2), 
(i,j = \pm)$, where i.e. $R_{+-}$ denotes the
coincidence count rate between the + labeled detector at apparatus
1 and the - labeled one at apparatus 2. (For further technical information we refer the reader to our 
full length
paper \cite{Bellfullength}.)

From these four coincidence count rates we compute the correlation function:
\beqa
&& E(\delta_1,\delta_2) := \\ \nonumber
&& \frac{R_{++}(\delta_1,\delta_2) - R_{+-}(\delta_1,\delta_2) - 
R_{-+}(\delta_1,\delta_2) + R_{--}(\delta_1,\delta_2)}
{R_{++}(\delta_1,\delta_2) + R_{+-}(\delta_1,\delta_2) +
R_{-+}(\delta_1,\delta_2) + R_{--}(\delta_1,\delta_2)}
\label{correlation}
\eeqa
and can determine the Bell parameter:
\begin{equation}
S = |E (d_1, d_2) + E (d_1, d_2') + E (d_1', d_2) -
E (d_1', d_2')| \leq 2,
\label{Bell}
\end{equation}
where $d_i, d_i'~(i=1,2)$ denote values of phases $\delta_i$. The above inequality, known as  
Bell-CHSH inequality \cite{CHSH}, is
satisfied by all local theories. Quantum mechanics predicts a maximal value
for the  Bell parameter $S=2\sqrt{2}$. 

Another type of Bell-inequality was given by Clauser and Horne \cite{CH} for an 
experiment with polarizers. A similar argument can 
be applied to experiments using interferometers:
if it is found experimentally that the single count rates are constant, and that 
$E(\delta_1,\delta_2)=E(\Delta)$ holds where $\Delta=(\delta_1+\delta_2)$ is 
the sum of the phases in both interferometers , then Eq.
\ref{Bell} reduces to
$S = |3E (\Delta) - E (3\Delta)|  \leq 2$. 
Beyond that, if it is found that the correlation coefficient E is described by a sinusoidal function of the 
form $E=Vcos(\Delta)$ with visibility V,
then the Bell parameter S becomes
$S = V 2\sqrt{2}$.
Hence, observing a visibility V greater than 
$V \geq \frac{1}{\sqrt{2}}\approx 0.707$
will in this case directly show that description of nature as provided by quantum mechanics 
is unreconcilable with the assumptions leading to the Bell inequalities.

In a first experiment, we changed the path length differences of both interferometers simultaneously, 
but at different
speeds. 
Comparing the correlation functions when both interferometers scan
in the same
direction, both in opposite directions and when only one is scanning, we
can confirm that
in a Franson-type interferometer the fringes can be described by a sinusoidal function and 
depend on the sum of the two
phases 
($\delta_1+\delta_2$). In addition, no phase dependent variation of the single count rates 
could be observed. Hence we can calculate the parameter S from the observed 
visibilities. In all cases we find values exceeding the limit given by the Bell-inequalities
by at least 9 standard 
deviations ($\sigma$).
The raw data for one of the best violations yield
$S_{raw}=(0.853\pm0.009)\cdot2\sqrt{2}$,
corresponding to a violation by 16 $\sigma$. Most
of the difference between this result and the theoretical prediction can be
attributed 
to accidental coincidences \cite{accidentals}. Indeed, from the measured single count rates (39.5 kHz) 
and the coincidence
window of (550$\pm$10) ps one can estimate the accidental coincidence rate to be 25.7$\pm0.5$ per 30
seconds (assuming that all events at both detectors are  uncorrelated). This
rate is in excellent agreement with the one we measured placing the coincidence window apart from the 
coincidence peak (26.4$\pm1.3$ per 30 sec). Subtracting the accidental coincidences, we
obtain 
$S_{net}=(0.955\pm0.01)\cdot2\sqrt{2}$, 
corresponding to a violation of the inequality by 24.8 $\sigma$. Since the visibility of 
the correlation function after subtracting the
accidentals is close to 1, 
one has to conclude that the distance does not affect the nonlocal
aspect of quantum mechanics, at least for distances up to 10 km \cite{problems}.

In a second experiment, we replaced one of the interferometers by 
two interferometers connected to the fiber from the source by a fiber
coupler (i.e.
a beam splitter). These two interferometers, however, used no circulators,
hence only
one detector per interferometer could be used. 
For this reason we can only measure two of the four coincidence count rates needed to 
calculate the correlation function (Eq. 1). To infer 
from the measured functions to the correlation function we thus have to assume the same 
symmetry between 
the coincidence functions as we 
found in the experiment described before. With this quite natural assumption, we can evaluate the 
correlation coefficients $E(d_1, d_2)$ and $E'(d_1', d_2)$ at the same time, 
hence for exactly the same setting $\delta_2$. Fig. 2 shows the correlation coefficients 
observed when 
changing the phase $\delta_2$ in the Bernex interferometer. We find again  
sinusoidal functions. Visibilities are about 78$\%$ without and about 96 $\%$ with 
subtraction of accidental coincidences. (The smaller raw visibility compared to the first experiment 
is due to 50 $\%$ additional losses in the coupler \cite{visibilities}.) 
We can 
now directly evaluate the value of the Bell parameter S (Eq. 2) from the correlation coefficients 
for two different values $d_2, d_2'$. For the indicated points we find $S_{raw}$ = 2.38 $\pm 0.16$ and 
$S_{net}$ = 2.92$\pm0.18$ leading to a violation of 2.4 respectively 5.1 standard deviations and 
confirming once again the quantum mechanical predictions.

Assuming that the passive
coupler
randomly selects which interferometer analyses the photon, this experiment can be
considered as involving truly random choices for the analyser settings as required to close the locality 
loophole \cite{locality}, at
least on
one side of the experiment. Since we find the same net 
visibility as in the first experiment, we can infer that the random choice at 
the beamsplitter does not change the result of the measurement.
One could argue that the choice is not really random, since the assumed
local hidden
variable could determine into which interferometer the photon is guided. Note
however, first, that
it is difficult to think of a better random number generator than a quantum one 
(based e.g. on a beam splitter as in our case), next
that if the hidden variable could determine a prefered interferometer, it could
equally well determine whether the photon is detected at all or remains undetected. 
This is the basis of the
detection loophole, an interesting possibility still open for local
theories
\cite{loophole}.

Another way to look at our experiments is quantum
cryptography based on entangled particles \cite{Ekert91}. 
The quantum bit error rate (QBER) \cite{QBER}
of this scheme
is related to the visibility $V$ before removal of the accidental coincidences: 
QBER=$\frac{1-V}{2}$. Note that subtracting the 
accidentals is impossible for quantum
cryptography, as there is no way to determine which coincidence counts are accidental and
which are due to a photon pair. From our measured raw
visibility of 85.2\% we infer a
QBER of 7.4\%. This is higher than the QBER obtained in experiments using weak pulses
\cite{PhysWorld}. Nevertheless, 
our result demonstrate that it is promising for practical implementation,
not so far from the schemes working with weak pulses. 
A fast switching in order to really exchange a key still has to be 
implemented.
This switching can be done either by a phase modulator or, as we did in our 
last experiment, by using a 
fiber coupler connected to two interferometers with appropriate phase differences. 
The advantage of the latter
setup is that no fast random generator and switching electronic is necessary. 
However, since the QBER increases with increasing losses, this setup would in our 
case be limited to around 10 km, a distance which is determined by the number of created photon pairs, 
overall losses and detector performance. 
A better way to do entanglement-based quantum cryptography 
would be to use a source employing nondegenerate phasematching in order to create correlated photons 
of different 
wavelengths, one at 1310 nm, the other one around 900 nm. 
This would allow to use more efficient and 
less noisy silicon photon counting modules to detect the photons of the lower wavelength.
To avoid the high transmission losses of photons of this 
wavelength in optical fibers, the interferometer(-s) measuring these photons could be placed next 
to the source.
First investigations show that quantum cryptography over tens of kilometers should be possible. 
It is interesting to note that besides ensuring the security of entanglement 
based quantum cryptography, the Bell inequality 
is even connected to the one qubit application of quantum cryptography: a quantum channel can be used
safely if and only if the noise in the channel is small enough to allow a violation
of Bell inequality \cite{EveBell}. 

As already mentioned in the introduction, no experiment up to date has been loophole free. 
Assuming that our results are not affected by the presence of these loopholes, 
this experiment demonstrates that energy-time entanglement
is robust enough to manifest itself in the violation of Bell inequality by
photons more than 10 km apart.
It opens also the door to several new possibilities: close the
locality
loophole, densecoding \cite{densecoding}, entanglement swapping \cite{entswap} and
quantum teleportation \cite{Qteleportation} at large distances as well as for entanglement 
based quantum cryptography. There is
also another interesting proposal: set the two analyser in motion such
that each analyser in his own inertial frames measures the photon pairs
first \cite{Suarez97}.

This work was supported by the Swiss FNRS and Priority Program for Optics,
by the European TMR network "The 
physics of Quantum Information", contr. no. ERBFM-RXCT960087, and by the
Fondation Odier. We like to thank G. Ribordy for help during the experiment as well as 
J. D. Gautier and O. Guinard for technical support.
The access to the telecommunication network was provided by Swisscom and the
circulators by JDS.



\begin{figure}[b]
\infig{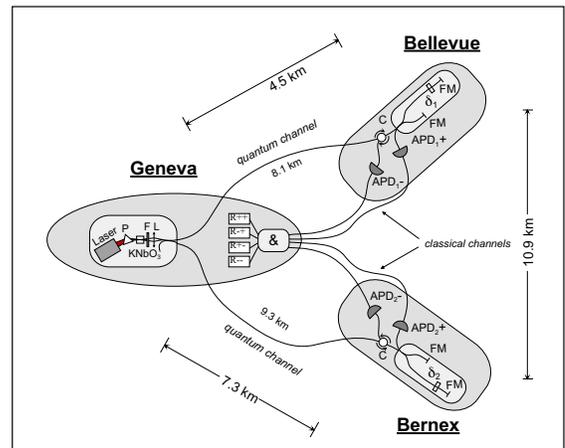}{0.85\columnwidth}                             
\caption{Setup for experiment 1. See text for detailed description} 
\label{fig1} 
\end{figure}

\begin{figure}[b]
\infig{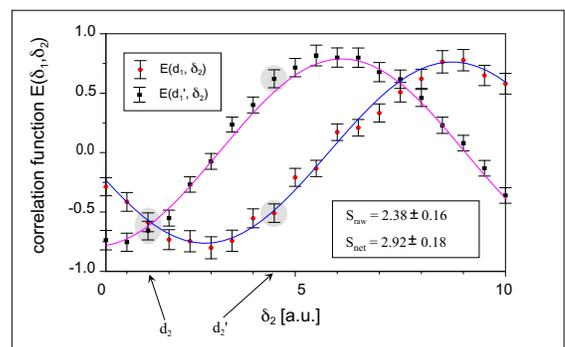}{0.85\columnwidth}                             
\caption{Result for experiment 2: The correlation functions $E(d_1, \delta_2)$ and $E(d_1', \delta_2)$
are plotted as a function of phase 
$\delta_2$. From the four indicated points one obtains $S_{raw}$=2.38 $\pm0.16$ and 
$S_{net}$=2.92$\pm0.18$, leading to a violation of the CHSH-Bell-inequality of 2.4 and 5.1 standard 
deviations, respectively.} 
\label{fig2} 
\end{figure}


\begin{references}
\bibitem{notthesame}This contrast with the experiment reported in \cite{Tapster94}
                where both anayzers where sitting next to the source, one connected with a
                4.3 km fiber on a spool, 
                while the other analyzer was directly connected to the source. Therefore the 
                physical distance between the two analyzers was of only a few meters.
\bibitem{1310} Let us emphasize that the choice of the wavelength is dictated by 
                the requirement of low fiber losses and the zero wavelength of chromatic dispersion.
\bibitem{accidentals}Accidental coincidences emerge from simultaneous 
                (i.e. within a 550 ps coincidence window) detection of uncorrelated photons. They are
                caused by simultaneous dark counts in two 
                detectors, by simultaneous detection of two photons belonging to different (noncorrelated)
                pairs being created at the same time,
                or by simultaneous detection of one photon of a pair and a dark count in the 
                other detector, 
                while the correlated photon was absorbed by the fiber.
\bibitem{problems}Besides determining the correlation functions, we made fits 
                of the underlying coincidence functions $R_{i,j}, (i,j=\pm1)$ as well \cite{Bellfullength}.
                We find them to 
                be in close agree with the values 
                for the correlation function. However, even if 
                the single count rates show almost 
                perfect symmetrie, we find a difference in the mean values 
                of the coincidence counts. Therefore the sum of 
                the coincidence rates of one detector with both detectors 
                on the other side (i,e. $R_{++}$ + $R_{+-}$) 
                is not constant. The origin of this difference is probably due to different
                afterpulse fractions or timejitter 
                of the detectors. However, if summing over all coincidence rates, we 
                always find the same value, confirming 
                that the size of the detected sample of photon pairs is constant.
\bibitem{QBER}The QBER is defined as the ratio of wrong bits to the total amount of 
                detected bits.
\end{references}
\end{document}